\documentclass[graybox]{svmult}

\usepackage{mathptmx}       
\usepackage{helvet}         
\usepackage{courier}        
\usepackage{type1cm}        
\usepackage{makeidx}         
\usepackage{graphicx}        
\usepackage{multicol}        
\usepackage[bottom]{footmisc}
\usepackage{url}

\makeindex             

\graphicspath{{./chapters/Sharma/}}

\begin{document}

\title*{Patterns of Linguistic Diffusion in Space and Time: The Case of Mazatec}

\titlerunning{Linguistic diffusion -- The case of Mazatec}

\author{Jean L\'eo L\'eonard, Els Heinsalu, Marco Patriarca, Kiran Sharma, Anirban Chakraborti}

\institute{
Jean L\'eo L\'eonard \at Paris-Sorbonne University, STIH, EA 4509, France.\\
\email{leonardjeanleo@gmail.com}\\
\\
E. Heinsalu and M. Patriarca \at NICPB--National Institute of Chemical Physics and Biophysics, R\"avala 10, 10143 Tallinn, Estonia.\\
\email{els.heinsalu@kbfi.ee; marco.patriarca@kbfi.ee}\\
\\
Kiran Sharma and Anirban Chakraborti \at SCIS--School of Computational \& Integrative Sciences, Jawaharlal Nehru University, New Delhi - 110067, India. \\
\email{kiran34_sit@jnu.ac.in; anirban@jnu.ac.in}
}

%
%
\maketitle


\abstract{In the framework of complexity theory, which provides a unified framework for natural and social sciences, we study the complex and interesting problem of the internal structure, similarities, and differences between the Mazatec dialects, an endangered Otomanguean language spoken in south-east Mexico.
The analysis is based on some databases, which are used to compute linguistic distances between the dialects.
The results are interpreted in the light of linguistics as well as statistical considerations and used to infer the history of the development of the observed pattern of diversity.
}

\section{Introduction}
\label{introduction}

Complexity theory is a major interdisciplinary paradigm, which provides a unified framework for natural and social sciences. At an operative level, it is based on a combined application of quantitative and qualitative methods at various phases of research, from observations to modeling and simulation to the interpretation of complex phenomena (Anderson 1972, Ross \& Arkin 2009).
Among the many applications, ranging from physics to biology and the social sciences, the study of language through the methods of complexity theory has become an attractive and promising field of research. In this contribution, we consider the complex and interesting case of the Mazatec dialects, an endangered Otomanguean language spoken in south-east Mexico by about 220,000 speakers (SSDSH 2011-16; Gudschinsky 1955, 1958).

\subsection{General Method}
\label{method}

Language dynamics represents a relevant branch of complexity theory, which  investigates  the classical problems arising in the study of language through novel approaches. Several methods have been imported directly from various scientific disciplines and used to model language from different points of view and at different levels of detail, which complete each other providing, all together, a new informative picture. Among these models and methods, one finds for instance:

(a) simple models addressing the language dynamics of population sizes at a macro- or meso-scopic scale, as in the ecological modeling \`{a} la Lotka-Volterra (Heinsalu, Patriarca \& L\'eonard 2014), which are able to tackle delicate issues such as the perceived status of languages (which directly affect the one-to-one language interaction between individuals) and describe other social features;

(b) nonlinear and stochastic dynamical models, reaction-diffusion equations, etc.,  which allow one to investigate at a meso-scopic level the most different issues and effects, related, e.g. to population dynamics, the spreading in space of linguistic feature on the underlying physical, economical and political geography (Patriarca \& Heinsalu 2009);

(c) individual-based models at the microscopic level, which are used to make numerical experiments to study languages along the perspective of language evolution (Steels 2011) and language competition, i.e., the dynamics of language use in multilingual communities (Sole, Corominas-Murtra \& Fortuny 2010, Stauffer  and Schulze 2005, Wichmann 2008, San Miguel \& al.2005). The latter topic is deeply linked to social interactions, thus the models used have direct connections with social sciences and social dynamics. In fact, linguistic features can be considered as cultural traits of a specific nature and their propagation can be modeled similarly to cultural spreading and opinion dynamics processes (Castellano, Fortunato, Loreto 2009; San Miguel, Eguiluz, Toral \& Klemm 2005).

\subsection{Plan of the work -- application to Mazatec dialects}
\label{plan}

The Mazatec dialects are localized in south-east Mexico. The approximate population of 220,000 speakers is characterized by a highly heterogeneous culture  and a locally diversified economic production landscape. The Mazatec dialects have become a classical topic in dialectology, due to the fact that they offer the typical highly complex panorama usually observed when studying cultural landscapes, in particular those characterizing endangered languages (SSDSH 2011-16; Gudschinsky 1955, 1958, 1959; Kirk 1966; Jamieson 1988, 1996; L\'eonard, dell'Aquila \& Gaillard-Corvaglia 2012; L\'eonard \& Kihm 2014).
This paper consists the analysis of the Mazatec dialects and in particular  their mutual linguistic distances, relying on previous and more recent databases and data analyses by various field-linguists. Such results will be reanalyzed and visualized using the tools of complex network theory, providing us with a measure and a picture of their homogeneity and heterogeneity . Different types of data will be considered, such as those related to the average linguistic Levenshtein distance between dialects (Heeringa \& Gooskens 2003; Bolognesi \& Heeringa 2002, Beijering, K. Gooskens C. \& Heeringa 2008) or those extracted by a direct comparison between speakers, i.e., based on the mutual intelligibility of dialects (Kirk 1970, Balev et. al 2016).
In Section \ref{ecology}, relying on the knowledge of the system (and in particular of the values of its main parameters) gained by the work carried out thus far (Kirk's comparative phonological database for interdialectal surveys and fieldwork). we will take into account external constraints such as the ecology of the settlement settings throughout the threefold layered system of  Lowlands, Midlands and Highlands, as well as the more recently superposed social and economic impact of postcolonial agro-industrial systems, such as coffee, cattle breeding and sugar-cane (all related, e.g., to the agricultural use of the land). 
In Section \ref{miniature}, the comparison between the picture suggested by the complex network analysis of the various data sets (overall sample of lexical categories versus a noun data base, restricted to phonological analysis) and other relevant aspects of the system under study will be carried out. This includes comparison of the linguistic networks with the underlying road networks, physical geography and economical geography. We will avoid \textit{materiality} such as ecological settings, to constructs like dialect areas, in order to account for the evolution of a very intricate diasystem, and end with a set of proposals for diasystemic geometry (a component of language dynamics) as a promising field for complexity theory.

\section{Language ecology}
\label{ecology}

\subsection{Ecological settings}
\label{settings}

Mazatec has resisted assimilation in the long term, thanks to its demographic weight (more than 200,000 speakers) and to emerging language engineering for literature and education through modern spelling conventions but it is still a very vulnerable language. The data collected in the ALMaz (A Linguistic Atlas of Mazatec; see L\'eonard \& al. 2012) support a pessimistic impression, also considering the collapse of the more recent agrarian systems of coffee crops and cooperatives, the consequences of the Miguel Alem\'an's dam in the 1950's, still to be seen [see Meneses Moreno 2014, Schwartz 2016] and a constant drive of migration to urban centres such as Tuxtepec, Tehuac\'an, Oaxaca, Puebla, M\'exico DF, or the USA.
The Mazatec area stands in the very centre of the Papaloapam Basin, benefiting from a smooth transition between the plain (e.g., Jalapa de Diaz) and the mountains,  West of the Miguel Alem\'an dam. This ecologically strategic position turned out to be fatal to the Mazatec Lowlands, partly drowned by the Miguel Alem\'an dam in the mid-50's, when the Rio Tonto, a powerful river connected to the Papaloapam mainstream, was controlled for the benefit of beverage and hydroelectric companies. Sugar cane also demands much water for crops. Patterns of cross-regional integration, which had quietly evolved since Olmec times (Killion \& Urcid, 2001) were disrupted in one of the few regions where native peasants (Mazatec and Chinantec mostly) worked their own microfunding.
Maps in Fig. \ref{fig1}-\ref{fig2} enumerate the Mazatec municipalities from Baja to Alta Mazateca (Lowlands and Highlands), providing an explicit view of the landscape: to the east, a plain half drowned by the dam (the Lowlands), to the west, a high Sierra mountain chain divided in the south by a canyon -- the Cuicatl\'an Canyon, with the Mazatec small town of Chiquihuitl\'an, famous for Jamieson's grammar and dictionary, published by the SIL in the late 80's and mid-90's (Jamieson, 1988, 1996).
Fig. \ref{fig1} (Left) provides an orographic and hydrographic map of the Mazateca area.
Fig. \ref{fig1} (Right) shows the distribution of Municipios over the Mazatec area -- the shape of the spots on the maps in Fig. \ref{fig1} (Left) and \ref{fig2}  hints at demographic size for each town, whereas Fig.~\ref{fig2} points out the \textit{ municipios} visited for the ALMaz since 2010 (in this map, only localities already surveyed by Paul Livingston Kirk are mentioned, showing how the ALMaz network is intended to be much larger than in previous dialectological studies, as Kirk 1966, Gudschinsky 1958, 1959). 
The Mazatec diasystem (Popolocan, Eastern Otomanguean) can be divided into two main zones: the Highlands and the Lowlands. Other subzones can be further distinguished, such as the Midlands (Jalapa de Diaz, Santo Domingo, San Pedro Ixcatl\'an) within the Lowlands, the Cuicatl\'an Canyon (Chiquihuitl\'an) and the Puebla area (see San Lorenzo data below). In short, main dialect subdivisions read as follows (slightly modified from L\'eonard \& Fulcrand 2016):

{\begin{center}
\textbf{(I) The Mazatec diasystem-- Dialects and sub-dialects}
\end{center} 

\begin{enumerate}
\item {\bf Highland complex}:\\
\textit{Central Highlands} (Huautla de Jim\'enez, Santa Maria Jiotes, San Miguel Huehuetl\'an)\\
\textit{Northwestern Highlands}\\
\indent -- Central Northwestern Highlands (San Pedro Ocopetatillo, San Jeronimo Tecoatl, San Lucas Zoquiapam, Santa Cruz Acatepec, San Antonio Eloxochitl\'an)\\
\indent -- Peripheral Northwestern Highlands (San Lorenzo Cuaunecuiltitla, Santa Ana Ateixtlahuaca, San Francisco Huehuetl\'an)

\item {\bf Lowland complex}:\\
\textit{Eastern Lowlands} (San Miguel Soyaltepec)\\
\textit{Central Lowlands} (San Pedro Ixcatl\'an)\\
\textit{Piedmont or Midlands} (Ayautla, San Felipe Jalapa de Diaz, Santo Domingo)

\item {\bf Periphery}:\\
\textit{South-Western Highlands} (Mazatl\'an Villa de Flores)\\
\textit{Cuicatl\'an Canyon} (Chiquihuitl\'an).\\
\end{enumerate}

It should be kept in mind that such a classification is not exhaustive but provides only a heuristic framework to observe variation.

%
\begin{figure}[ht!]
\centering
\includegraphics[width=0.48\linewidth]{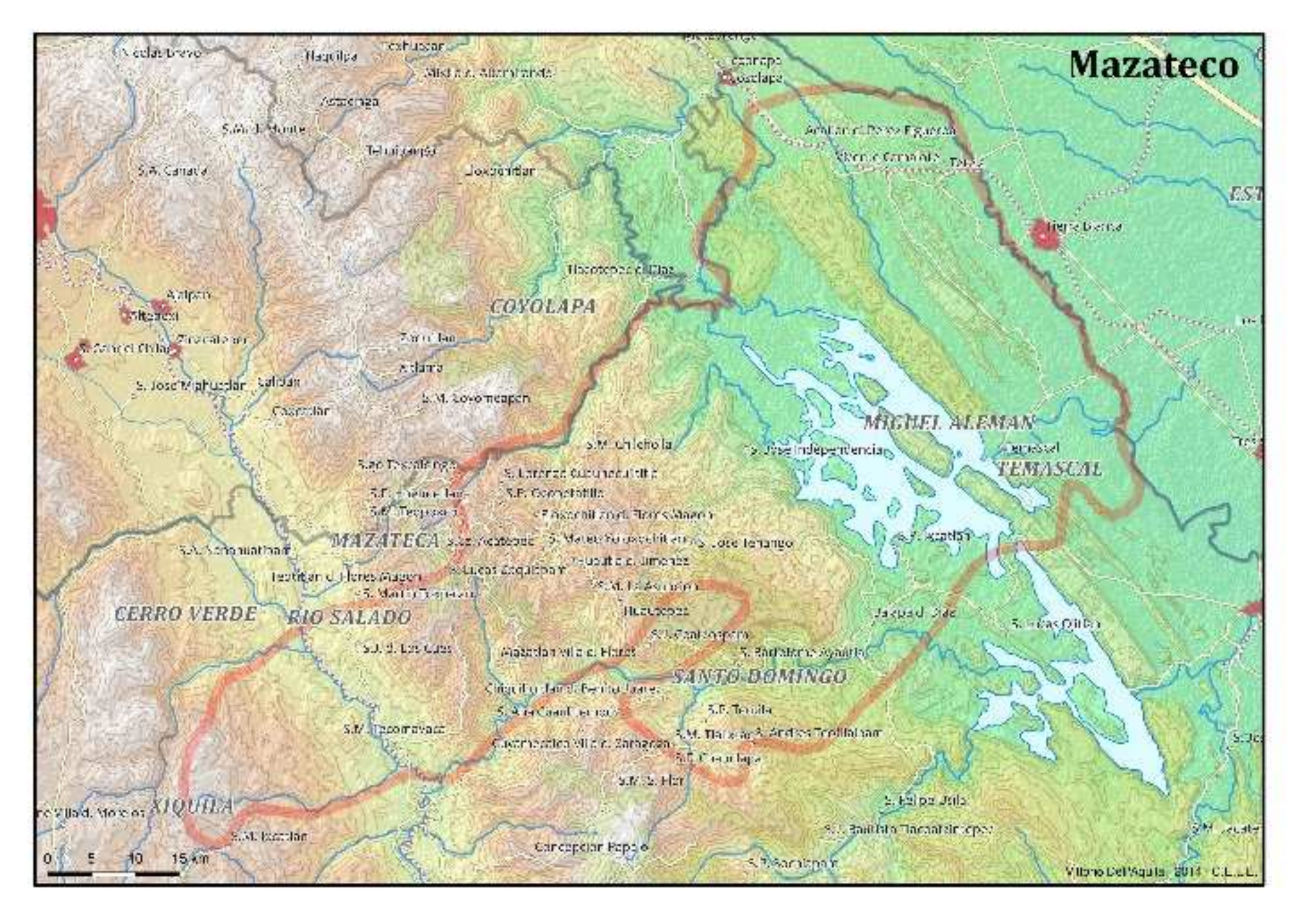}
\includegraphics[width=0.48\linewidth]{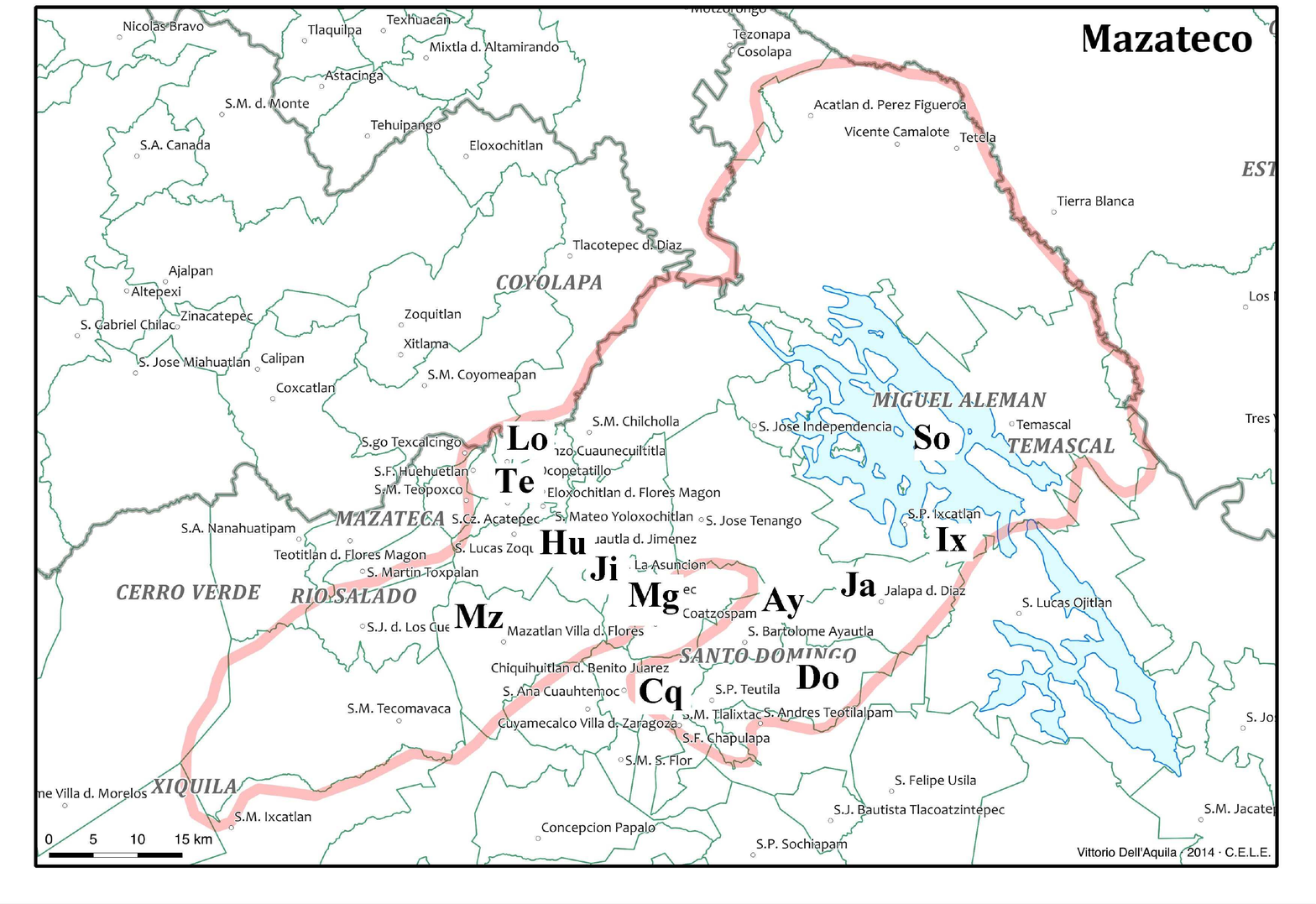}
\caption{The Mazatec dialect network (localities surveyed in Kirk 1966). Maps: CELE (Vittorio dell'Aquila 2014).}
\label{fig1}
\end{figure}

%

The spots on the map  in  Fig. \ref{fig2}, cluster into significant sub-areas. Behind the dam stands San Miguel Soyaltepec, a very important centre from ancient times, which was probably connected through the plains to the coastal zone of the Papaloapam Basin. From the size of the spots in Fig.\ref{fig2}, revealing the  demographic weight, we can state that it is still the biggest urban centre in the Mazatec lands.

%
\begin{figure}[ht!]
\centering
\includegraphics[width=12cm]{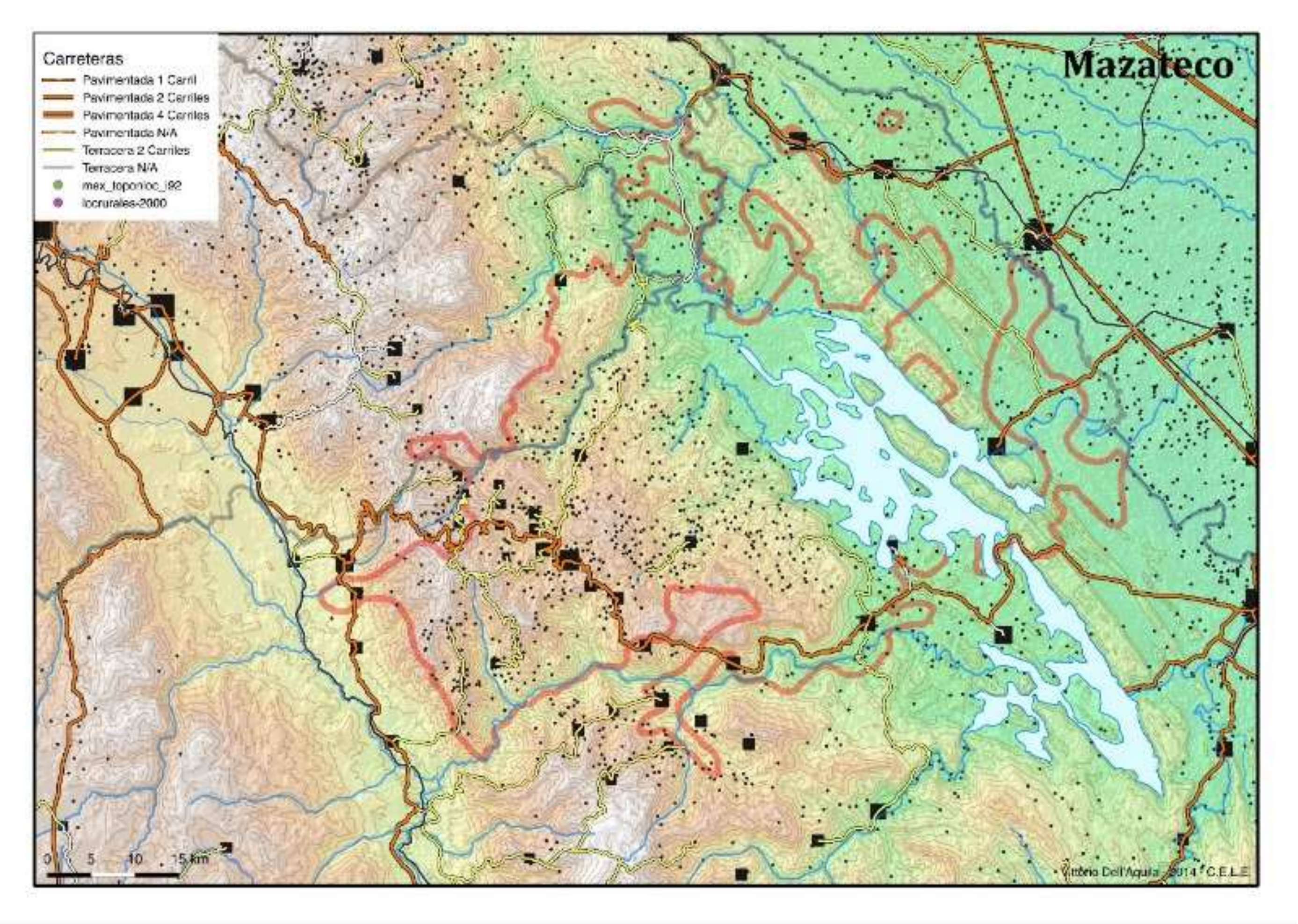}
\caption{Communal aggregates in the Mazatec area. Map: CELE (Vittorio dell'Aquila 2014). Official census data (2002).}
\label{fig2}
\end{figure}

The town of Acatl\'an,  north of Soyaltepec, is more Spanish speaking than Soyaltepec. Inhabitants of the archipelago inside the artificial lake -- within the huge pool created by the dam -- use the same variety as in San Miguel Soyaltepec, as do the new settlements, such as Nuevo Pescadito de Abajo Segundo, in the South. A dialect network probably as intricate as that of the North-West Highlands (around San Jerónimo Tecoatl) probably existed before the flooding of the microfundio agrarian society of the Lowlands. Most of these dialects merged into mixed dialects, apparently under the strong influence of the Soyaltepec koin\'e (we use this term as ``local speech standard'', i.e. pointing at an oral, more than a written koin\'e, though nowadays a Soyaltepec written koin\'e does exist, strongly supported by local poets and school teachers). This first segment of the Mazatec world makes up the San Miguel Soyaltepec Lowlands segment: a resilient area, with a strong urban constellation going from the newly built Temascal to the industrial town of Tuxtepec, with strong local dialect intercourse and mingling, in a region whose agrarian structure has been  drowned by a pharaonic dam project sixty years ago. The consequences of this dramatic redistribution of agrarian resources and property, and of the displacement of over 220,000 peasants, are still to be seen. Linguistically, this event partially enhanced acculturation and assimilation to Spanish under the influence of urban centres such as SM Soyaltepec, but most of all, Temascal, Acatl\'an, and Tuxtepec.
The second area, joning from Lowlands to Highlands, covers the western shores of the Miguel Alem\'an lake, as a twofold stripe, from S. M. Chilchotla and San Jos\'e Independencia (Midlands) to San Pedro Ixcatl\'an (Western Lowlands), in the continuity of the plain or the valley, where  the important urban centre of Jalapa de Diaz is located. This Midland-Lowland region displays a whole range of small urban centres, dominated by sugar-cane and herding (the agrarian couple ca\~na y ganado). Though we should consider Jalapa de Diaz as a subarea of its own, because of its size and its links with other regions, such as the Highlands (Huautla) and the so called Ca\~nada or Canyon (Chiquihuitl\'an and beyond), we may lump both subareas as the Western Plain. 
The Highlands qualify as the third main area, after the subdivisions of the Lowlands into the SM LL and the Western Plain. In turns, it divides into two subareas: central, with Huautla, and the Western Highlands -- a dense network of small urban centres such as San Lucas, San Jerónimo Tecoatl, San Lorenzo, San Francisco Huhuetl\'an, and San Pedro.
We will call the fourth complex ``the Ca\~nada Connection'', where the most conspicuous urban centre is Mazatl\'an de Flores, on the periphery of the Canyon, and Chiquihuitl\'an. This is a region of intense language contacts: from Chiquihuitl\'an downhill through the Canyon, Cuicateco, a Mixtecan language is spoken. Nowadays, the zone seems to have fallen into the hands of the Narcos, and the road to Chiquihuitl\'an is no longer an easy to trip from Jalapa de Diaz, as the ALMaz staff has experienced in recent years. The dialect of a spot such as Santa Maria Tecomavaca, on the western plateau, has scarcely been documented up to now, though it is not so far from neighbouring centres such as Mazatl\'an or Teotitl\'an del Camino. Though, it forms a subarea on its own in the Canyon region, because of the low rate of Mazatec speakers as compared to the central area of the Mazatec world, and its location on the plateau, with a tropism outward of the Mazatec area (towards Teotitl\'an del Camino, Tehuac\'an, etc.). Strikingly enough, the variety spoken in this peripheral area has more to do with the Northwestern Highlands dialects than with the neighboring Mazatl\'an area, pointing at strong resettlement dynamics throughout the Mazatec area, far beyond the state of the art knowledge of these phenomena. To us, the main reason lies in the way the coffee economy drained people from the poorest regions of the Midland Outer Belt (Santa Maria Chilchotla, San Mateo Yoloxochitl\'an), towards the Teotitl\'an del Camino urban centre, where coffee used to be sold to merchants. Though, the San Juan de los C\'uest Santa Maria Tecomavaca still makes up an original dialect of its own, as several varieties apparently migrated there, from the early 19th to the end of the 20th Century.

The agrarian ecology of these subzones appears in Fig.~\ref{fig3} (Left).
Next, we will deal with socio-linguistic ecology, giving a few hints about linguistic vitality.

\subsection{Sociolinguistics: vitality zones}
\label{zones}

Areas and subareas can also be defined by the sole criterion of the rate of speakers,
as in Fig.~\ref{fig3} (Left), in territories considered as traditionally Mazatec. At first sight, we can see that the core of the Mazatec area still uses the language intensively (H index), whereas the periphery does not (L on the Eastern shore of the dam and in the Canyon. Two pockets have medium scores: ml at San Juan de los Cúes and mh at Chiquihuitl\'an.

%
\begin{figure}[ht!]
\centering
\includegraphics[width=0.48\linewidth, height=0.35\linewidth]{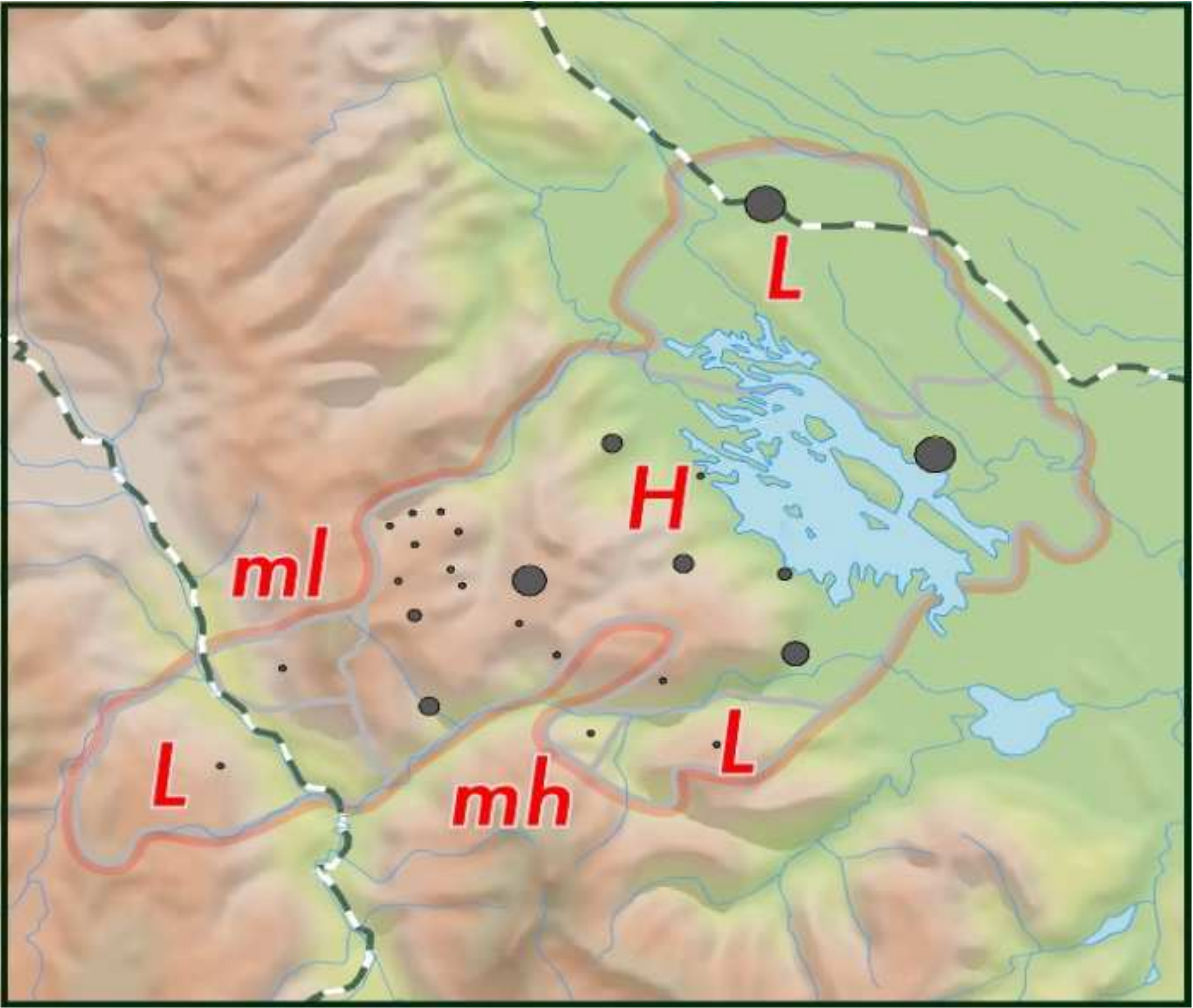}
\includegraphics[width=0.48\linewidth, height=0.36\linewidth]{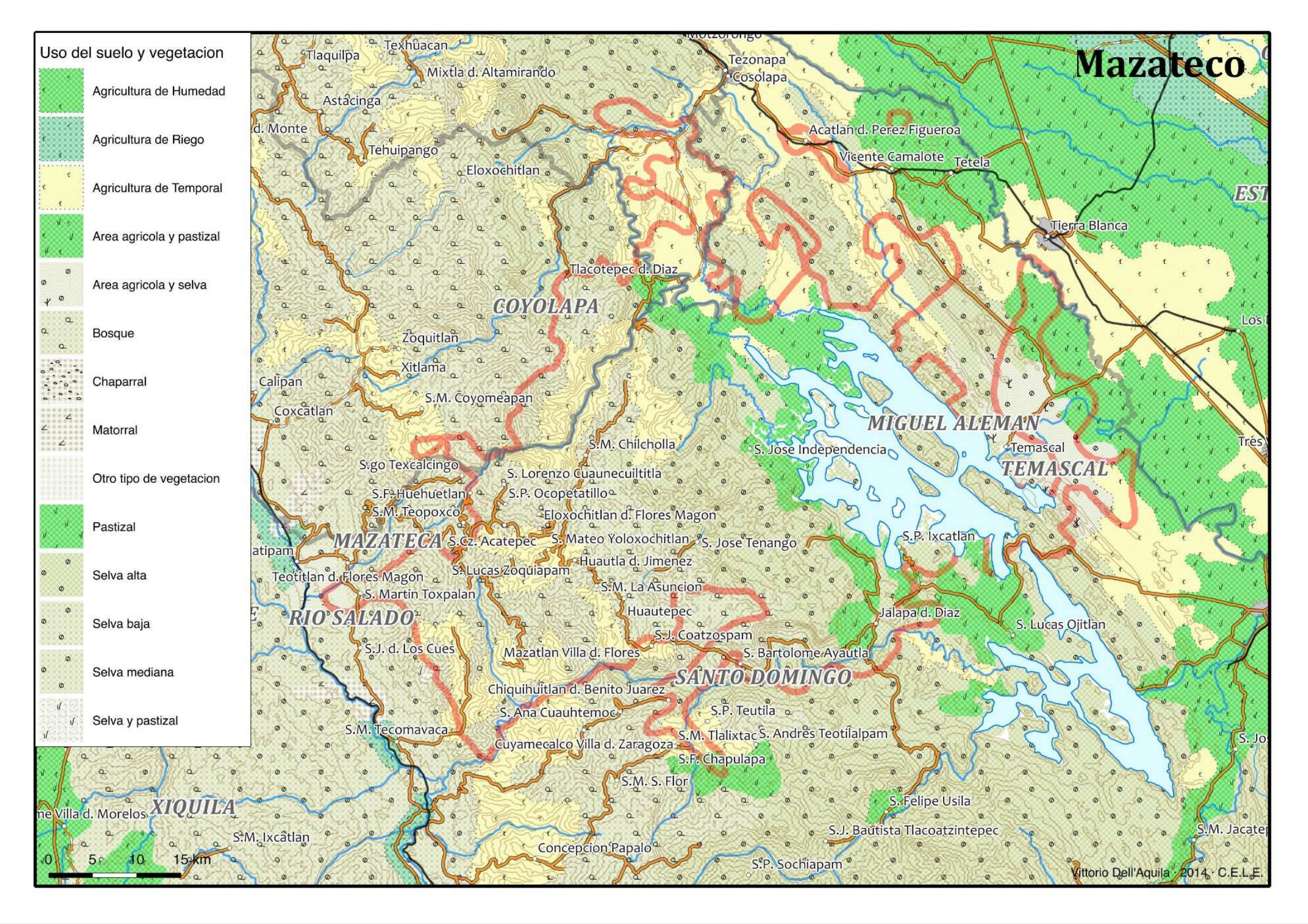}
\caption{(Left) Urban centers and degrees of vitality of Mazatec (Leonard \& dell' Aquila 2014). The rate of speakers is defined as: H = High rate of Mazatec speakers (over 75 \%),
mh = mid-high value, i.e. 50-75\% of the population speaking Mazatec, ml = mid-low density of speakers, i.e. 25-50\%, L = low density, i.e. 0-25\%.
(Right) Ecological and agrarian zones in the Mazatec area (CELE (Vittorio dell'Aquila 2014). }
\label{fig3}
\end{figure}

%

\section{Dialect dynamics: a study in miniature}
\label{miniature}

The title of this section takes over the subtitle of a seminal paper on Mazatec ethnohistory (Sarah Gudschinsky 1958), in which Gudschinsky claimed that geolinguistics provided reliable clues to the Proto-Mazatec evolution into five main dialects, through seven periods (Fig. \ref{fig3} (Right)): 

\begin{center}
\textbf{(II) Gudschinsky's 1958 dialect dynamics model}
\end{center} 

\begin{enumerate}
\item {homogeneity};

\item {slip according to alternating *a and *u};

\item {emergence of a Lowlands dialect, to which Mazatl\'an (MZ) and San Miguel Huautepec (MG) still belonged -- whereas the former is nowadays a peripheral Highlands dialect, the latter strongly clusters with Huautla (HU), in the Central Highlands area};

\item {the Valley dialect emerges (Jalapa, i.e., JA) and differs from MG, then the Southern valley dialects split from a Northern one, while `foreign domination' (Mixtec) takes hold of the region};

\item {the Highlands dialect emerges, and attracts MZ to its circle of influence, roughly during the period 1300 to 1456; two kingdoms compete, in the Highlands and the Lowlands respectively, (F) Western Highlands, MG and Norther Lowlands dialects differ, and Aztec rule takes hold}. 
\end{enumerate}

A more cautious model without so many details on the Mixtec and Aztec hegemonies was proposed previously by the same author (Gudschinsky 1955) and describes five differentiation periods (or phases):

{\begin{center}
\textbf{(III) Gudschinsky's 1955 dialect dynamics model}
\end{center} 

\begin{enumerate}
\item {Homogeneity, followed by the rise of HU and JA}.

\item {Emergence of a transitional buffer zone between HU \& JA}.

\item {(a) The lowland zone splits in two, with the emerging variety of IX.

(b) Both HU and IX areas diversify: SMt (San Mateo) emerges in the highlands, whereas SO splits from IX. In the buffer zone, MG  also emerges. Flows of lexicon and variables still pass from the Lowlands to the Highlands}. 

\item {Further and more clear-cut differentiation between IX and SO, in the Lowlands}.

\item {Consolidation of the six dialects: sharper frontiers}.
\end{enumerate}

In the next section, where the Levenshtein distance (LD) is applied to Kirk's data on twelve varieties (Kirk 1966) for surveying dialect dynamics, Gudschinsky's models as summarised in (II) and (III) above are very useful to interpret the results and suggest a better overall agreement with Gudschinsky's model (III) -- rather than with (II).

\subsection{Levenshtein distances (LD)}
\label{LDdef}

The LD is used to estimate an average linguistic distance between each pair of dialects from the set of the LDs between variants of the same nouns. 
The LD $L(a,b)$ is a basic measure of the level of difference between two strings $a$ and $b$, defined as the minimum number of operations (represented by insertions, deletions, or editions) needed to turn $a$ into $b$ or vice versa.
For instance, given $a$ = ``thia'' (``arm'', AY) and $b$ = ``t{\c s}ha'' (``arm'', JI), the LD between these two variants of ``arm'' is $L(a,b) = 2$, corresponding to the two changes h $\to$ {\c s} and i $\to$ h needed to turn one string into the other.
The  LD  $L(a,b)$ has the merit to be simple in definition and use.
Its simplicity, however, also represents its limit, due to its independence of the type of the actual operations (whether insertions, deletions, or editions), the number and type of characters changed (e.g., vowels or consonants), and of the order in which they are changed.

We represent two noun variants in dialect $i$ and dialect $j$ of the same semantic meaning, labeled $k$,  as $a_{i,k}$ and  $a_{j,k}$.
Namely, the locations of dialects are labelled by the index $i$ (or $j$), running from $i = 1$ ( $j = 1$ ) to the total number of locations $i = NL$ ( $j = NL$ ), while the label $k$ runs over all the $M_{i,j}$ pairs of nouns $a_{i,k}$ and  $a_{j,k}$ in  dialects $i$ and $j$ with a common semantic meaning, $k = 1, \dots , M_{i,j}$.
For a fixed pair of dialects $i$ and $j$ the corresponding LD $L_{i,j}^k = L(a_{i,k}, a_{j,k})$ are computed for all the variants $k$ available.
The set of  LD thus obtained are then used to compute the average (final)  LD $L_{i,j}$  between dialects $i$ and $j$,

\begin{equation}
L_{i,j} = \frac{1}{M_{i,j}} \sum_{k=1}^{M_{i,j}} {N_{i,j}}  L_{i,j}^k \, .
\end{equation}

Notice that this represents  a simple arithmetic average, meaning that all the distances are considered to have equivalent statistical weights. Repeating this calculation for all pairs of dialects ( $i$, $j$ ) allows to construct the ``Levenshtein matrix'', whose elements are all the average LDs $L_{i,j}$ defined above.
The Levenshtein matrix for the twelve Mazatec dialects studied is visualized in the Table.~\ref{table1} (for $NL  = 12$ locations, there are $NL(NL - 1)/2 = 66$ such distances).

\begin{table}[ht]
\centering

\begin{tabular}{|l|l|l|l|l|l|l|l|l|l|l|l|l|}
\hline
\#          & \textbf{AY} & \textbf{CQ} & \textbf{DO} & \textbf{HU} & \textbf{IX} & \textbf{JA} & \textbf{JI} & \textbf{LO} & \textbf{MG} & \textbf{MZ} & \textbf{SO} & \textbf{TE} \\ \hline
\textbf{AY} &             & 0.28        & 0.20        & 0.32        & 0.21        & 0.24        & 0.30        & 0.52        & 0.29        & 0.27        & 0.24        & 0.29        \\ \hline
\textbf{CQ} & 0.28        &             & 0.30        & 0.38        & 0.30        & 0.33        & 0.37        & 0.54        & 0.34        & 0.35        & 0.30        & 0.34        \\ \hline
\textbf{DO} & 0.20        & 0.30        &             & 0.33        & 0.19        & 0.11        & 0.33        & 0.54        & 0.27        & 0.26        & 0.24        & 0.28        \\ \hline
\textbf{HU} & 0.32        & 0.38        & 0.33        &             & 0.32        & 0.30        & 0.21        & 0.53        & 0.25        & 0.30        & 0.24        & 0.33        \\ \hline
\textbf{IX} & 0.21        & 0.30        & 0.19        & 0.32        &             & 0.22        & 0.31        & 0.53        & 0.29        & 0.27        & 0.24        & 0.25        \\ \hline
\textbf{JA} & 0.24        & 0.33        & 0.11        & 0.30        & 0.22        &             & 0.32        & 0.55        & 0.28        & 0.28        & 0.25        & 0.238       \\ \hline
\textbf{JI} & 0.30        & 0.37        & 0.33        & 0.21        & 0.31        & 0.32        &             & 0.55        & 0.33        & 0.28        & 0.24        & 0.28        \\ \hline
\textbf{LO} & 0.52        & 0.54        & 0.54        & 0.53        & 0.53        & 0.55        & 0.55        &             & 0.55        & 0.33        & 0.50        & 0.50        \\ \hline
\textbf{MG} & 0.29        & 0.34        & 0.27        & 0.25        & 0.29        & 0.28        & 0.33        & 0.55        &             & 0.25        & 0.24        & 0.31        \\ \hline
\textbf{MZ} & 0.27        & 0.35        & 0.26        & 0.30        & 0.27        & 0.28        & 0.28        & 0.33        & 0.25        &             & 0.22        & 0.29        \\ \hline
\textbf{SO} & 0.24        & 0.30        & 0.24        & 0.24        & 0.24        & 0.25        & 0.24        & 0.50        & 0.25        & 0.22        &             & 0.26        \\ \hline
\textbf{TE} & 0.29        & 0.34        & 0.28        & 0.33        & 0.25        & 0.28        & 0.28        & 0.50        & 0.31        & 0.29        & 0.26        &             \\ \hline
\end{tabular}
\caption{A Matrix of LD for 12 Maztec dialects, 117 cognates.  
	(source: data from Kirk 1966, data processing: CELE, Vittorio dell'Aquila 2014).}
\label{table1}
\end{table}

\subsection{An overall sample for LD}
\label{sample}

In this section, dialectological data from Kirk (1966) will be measured according to LD (see the chapter on Basque geolinguistics for methodological details). As this algorithm measures and ponders distance between dialects synchronically, most of the results rely upon phonological and morphological patterns. Etyma are not used, contrary to a phylogenetic approach.
We will thus consider these results as highlighting ontological distances and complexity between dialects (e.g. the most complex dialect here is LO, in the Poblano area, in the NW outskirts of the Mazatec dialect network. 

It is useful to study the network as a function of a threshold $d$.
To this aim, we first normalize all the LDs by dividing them by  the largest LD found in the system, so that all the LD values are in the interval (0,1). Then the value $d = 0$ corresponds to perfectly equivalent dialects, while the value $d = 1$ to the farthest couple(s) of dialects.
The method consists in setting a threshold on the LDs, i.e., plotting two dialect nodes $i$ and $j$ only if their LD is such that 

\begin{equation}
L_{i,j} < d \, .
\end{equation}

At $d = 0$, no link is shown because no dialect is perfectly equal to another dialect.
When gradually increasing $d$, then some dialect nodes become connected producing a linguistic network.
At the maximum value $d = 1$, all the dialect nodes appear and are connected to each other.
However, not all link strengths are equal.
A useful way to plot the network is to make links between nodes thicker, if the corresponding LD is smaller, so that they provide an intuitive visual idea of the strength of the linguistic link. We plot the networks for different threshold values $d=0.22, ..., 0.29$, as shown in Fig. \ref{fig4}.

%
%


%
\begin{figure}[ht!]
\centering
\includegraphics[width=0.48\linewidth]{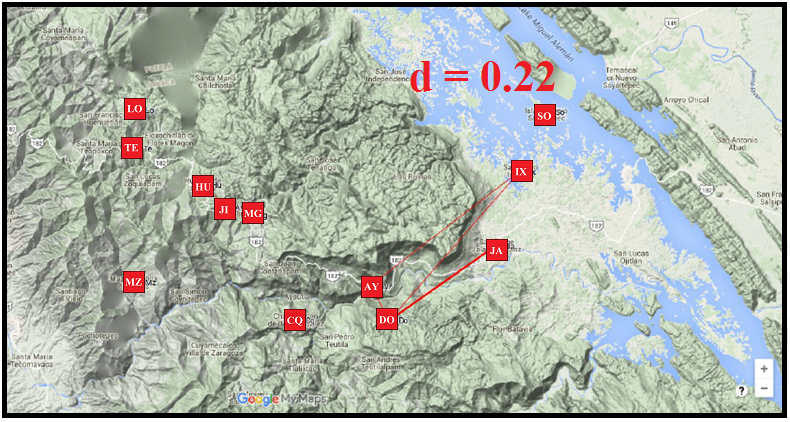}
\includegraphics[width=0.48\linewidth]{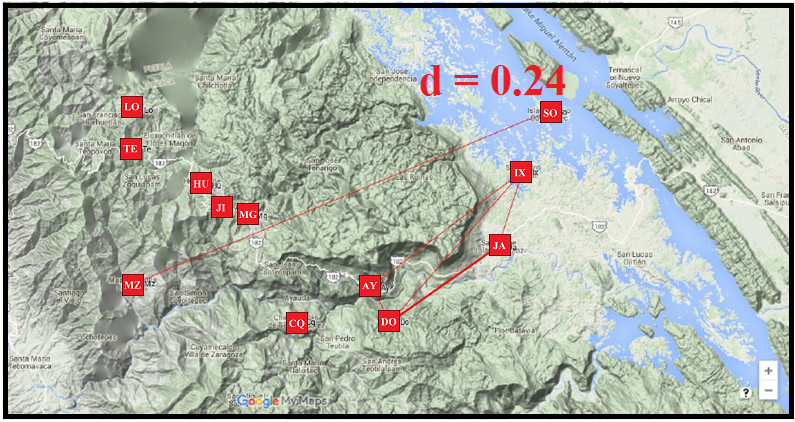}
\includegraphics[width=0.48\linewidth]{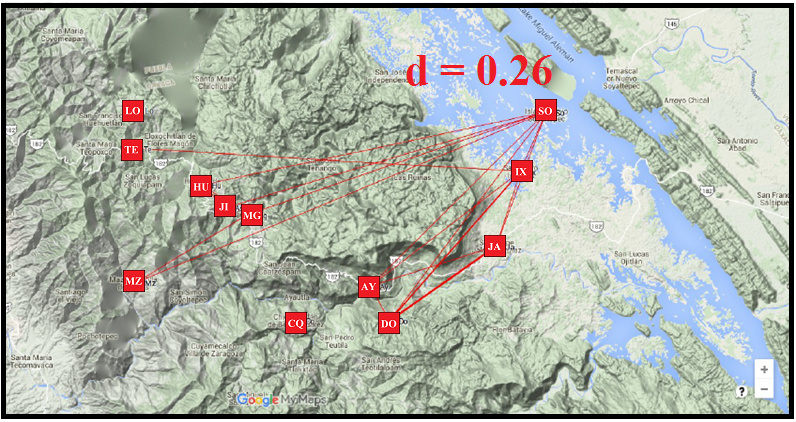}
\includegraphics[width=0.48\linewidth]{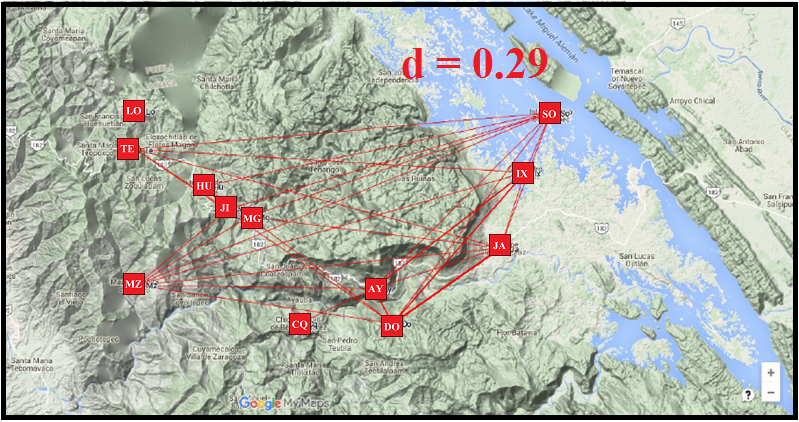}
\caption{Dialect network with threshold $d = 0.22$ (Upper Left),$ 0.24$ (Upper Right),$ 0.26$ (Lower Left) and $0.29$ (Lower Right).}
\label{fig4}
\end{figure}

Thus, Fig. \ref{fig4} (Upper Left) shows threshold $d = 0.22$ with the choreme (a kernel area, see Goebl 1998: 555).
The bolder line uniting JA and DO points at a dialect of its own, whereas the finer line, between DO and IX, resorts to a less organic structural relation, yet rather strong -- i.e., a chain, between this basic choreme [JA-DO] with the more autonomous and powerful Lowlands dialect of San Pedro Icxatl\'an. Another choreme is shown in the Highlands: HU and  JI, whereas the inner cohesion within the [IX[DO-JA]] chain is confirmed.
This [HU-JI] choreme will soon be connected to the most peripheral dialect, in the Eastern Lowlands (SO), and remains yet unconnected to close neighbors like MG or TE.
As we  soon shall see, these two choremes now available will soon raise their interconnectivity in the dialect network, enhancing patterns of resilience of a previous feature pool(see Mcfuene 2001,2012,2013) consistency in the valley.


In Fig. \ref{fig4} (Upper Right), with $d = 0.24$, a complex communal aggregate [MZ-SO], [HU-JI], [IX[DO-JA]] emerges. The pattern now points at two clusters [HU-JI], [IX[DO-JA]] and one far distant chain [MZ-SO].
As a matter of fact, all these patterns confirm Gudschinsky's model (1955), initially elaborated out of lexicostatistics.


In Fig. \ref{fig4} (Lower Left), with  $d = 0.26$, the overall picture becomes clearer, and goes far beyond Gudschinsky's expectations, in terms of fine grained representation of the intricacy of the diasystem; namely, we have a whole complex network with clear-cut communal aggregates: a [TE[SO[IX]]] chain, a [HU-JI-MG[SO]] chain, a macro-chain connecting in a most intricate way MZ with the [IX-DO-JA] chain, through AY and MG, working as areal pivots in the Midland and the Highlands respectively. The most peripheral varieties are LO in the Northwestern fringe, and CQ, in the Southwestern border of the Mazatec area. Interestingly enough, these spots are not connected yet in this phase,forming as what we can call ``default areas'' or ``default spots'', i.e. strongly divergent varieties, which do not correlate tightly enough with the rest of the network to highlight deep geolinguistic structures. Of course, one can cluster these erratic varieties, when elevating the threshold of divergence.


Fig. \ref{fig4} (Upper Left), with  $d = 0.29$, shows how CQ does correlate with already available clusters -- namely, with AY. Nevertheless, AY and CQ strongly differ in all respects, as our own fieldwork gave us evidence recently. The reason why CQ converges somewhat to AY is more due to the transitional status of AY, between the Highlands and the Lowlands,rather than to structural heritage, although indeed, these two variants can be seen as geographical neighbors.


The same could be said of LO, as compared to TE: the former finally connects to the latter in a nearest-neighbor graph, as shown in Fig. \ref{fig5}  (obtained by joining each dialect node only to the one from which it has the shortest LD) although the structural discrepancy is conspicuous. Indeed, LO proceeds from the same historical matrix as TE: the San Antonio Eloxochitl\'an dialect -- not surveyed by Paul Livingston Kirk, but from where we were able to elicit phonological and morphological data in 2011. This nearest-neighbor graph below provides a handy overall picture of the Mazatec dialect network, on the basis of the LD processing of our 117 cognates: it clearly highlights the far reaching interconnection of Highlands dialects with Lowlands dialects, with macro-chains [TE[IX]], [MZ[SO]] and the intricate cross-areal (i.e. Highlands/Lowlands) cluster [HU-JI-MG[SO]]. Lower range clusters, such as [AY[CQ[DO]]], and choremes, such as [DO-JA] and [HU-JI], as seen previously at stage $d = 0.22$, are also available in this map.

%
\begin{figure}[ht!]
\centering
\includegraphics[width=\linewidth]{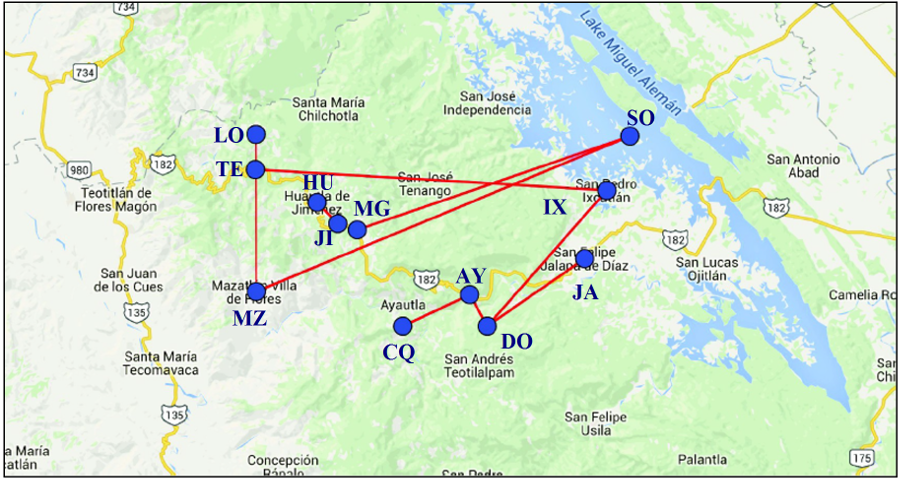}
\caption{Nearest neighbor network based on the LD distances of 117 cognates, based on the data of (Kirk 1966).}
\label{fig5}
\end{figure}

Considering Gudschinsky's model of dialect dynamics (III) above, one can now check to what extent its predictions were right. As a matter of fact, her claim (1) (homogeneity, followed by  the rise of HU and JA) is confirmed by phase $d = 0.22$, which clearly enhances the emergence of two choremes--high and low: [HU-JI] vs. [DO-JA]. 
Gudschinsky's period (2) entails the emergence of a transitional buffer zone between HU \& JA. This claim is strongly supported, but also enriched by phases $d = 0.24$ and $d = 0.26$: not only does HU cluster  with JI and MG, but AY also clusters with the IX and JA-DO chain. In turn, all these aggregates connect with Lowlands varieties, pointing at the formation of Highlands varieties as a by-product of Lowlands dialect diversification. The ambivalent structural status of MZ, standing far west in the Highlands, though connecting far into the East with SO, and even to IX, through the buffer area of AY, hypothesised by Gudschinsky in both models (II) and (III), is strongly confirmed too. Gudschinsky's Periods (3a-b), implying  the split of the Lowlands dialect in two (JA vs. IX) on the one hand (3a),  and on the other hand  the inner split of the Highlands (i.e. 3b:  HU vs. TE, standing for Gudschinsky's SMt , in this dialect network according to Kirk's data) are also confirmed by step $d = 0.29$, as these slots in the graph become more densely interactive with the rest of the dialect network. Though, results here display much more detail on general connectivity than in models in (II) and (III).
The period (4), with further and more clear-cut differentiation between IX and SO, in the Lowlands, is also confirmed by far reaching patterns of connectivity of SO with TE, HU, MZ in the highlands and AY in the Midlands. 
Results from these 117 cognates (see L\'eonard 2016: 77-79 for a complete list of items) are not simply congruent with Gudschinsky's hypothesis on dialect dynamics, as summed up in (II) and (III): they provide much more information about the hierarchization and intricacy of differentiation within the Mazatec dialect network. Moreover, they enhance the status and interplay of such (dia)systemic categories as choremes, chains, macro-chains and pivots or buffer zones. They also clearly point at a level of diasystemic organization which supersedes the Stammbaum and the chain level of organization: distant ties, either out of retention, or as an endemic effect of a feature pool (Mufene 2001, 2012, 2013) of traits inherited from the Lowlands dialects, which carried on mingling together long after the splitting of the main Highlands and Lowlands dialects. For example, many morphological facts point at an inherited stock of inflectional mechanisms in the Lowland dialects and peripheral Northwestern dialects such as LO (in Kirk's data) and San Antonio Eloxochit\'an (ALMaz data). The link between TE and IX in Fig.~\ref{fig5} confirms this trend -- whereas the link between HU and SO or MZ and SO may rely more on mere retention, and to an older layer of structural continuity. 
The sample processed here covered all lexical classes of the Mazatec lexicon, for a set of 117 cognates, from Kirk 1966: verbs, nouns, pronouns, adjectives, adverbs, etc. The results do provide a useful overall picture, but we  still suspect this sample to be too heterogeneous, and to blur finer grained patterns of differentiation within the lexicon and grammar. Verbs are especially tricky in Mazatec (Leonard \& Kihm 2014, Leonard \& Fucrand 2016), and bias may be induced by elicitation, for instance when the linguist asks for a verb in neutral aspect (equivalent to present tense) and may get an answer in the incompletive (future tense) or completive (past tense), or the progressive aspect, according to pragmatic factors (e.g., verbs such as `die' can hardly be conjugated in the present, as `he dies', and informants are prone to provide completive or incompletive forms, as `he died (recently)' or `he'll (soon) die'. Nouns in Mazatec are far less inflected than verbs -- only inalienable nouns, such as body parts and some kinship terms have fusional inflection (see Pike 1948: 103-106). The subset of nouns in the Kirk database, therefore, is more likely to provide abundant and much more reliable forms to implement the LD than a sample of all lexical categories.

\subsection{A restricted sample for LD}
\label{restricted}

Although this paper aims at modeling dialect dynamics rather than at providing a description of the language, some data may be useful at this point of the argumentation, in order to get a glimpse at word structure in Mazatec, and related processes on which the LD distance may apply.
\begin{table}[ht]
\centering
\begin{tabular}{|l|l|l|l|l|l|l|l|l|l|l|l|l|}
\hline
\#          & \textbf{AY} & \textbf{CQ} & \textbf{DO} & \textbf{HU} & \textbf{IX} & \textbf{JA} & \textbf{JI} & \textbf{LO} & \textbf{MG} & \textbf{MZ} & \textbf{SO} & \textbf{TE} \\ \hline
\textbf{AY} &        & 0.632       & 0.629       & 0.668       & 0.606       & 0.607       & 0.636       & 0.981       & 0.562       & 0.573       & 0.582       & 0.708       \\ \hline
\textbf{CQ} & 0.632       &        & 0.717       & 0.703       & 0.666       & 0.704       & 0.589       & 0.978       & 0.627       & 0.645       & 0.636       & 0.688       \\ \hline
\textbf{DO} & 0.629       & 0.717       &        & 0.689       & 0.585       & 0.334       & 0.643       & 1.000       & 0.608       & 0.639       & 0.620       & 0.703       \\ \hline
\textbf{HU} & 0.668       & 0.703       & 0.689       &        & 0.593       & 0.655       & 0.346       & 0.897       & 0.402       & 0.481       & 0.519       & 0.550       \\ \hline
\textbf{IX} & 0.606       & 0.666       & 0.585       & 0.593       &   & 0.599       & 0.616       & 0.937       & 0.574       & 0.639       & 0.519       & 0.586       \\ \hline
\textbf{JA} & 0.607       & 0.704       & 0.334       & 0.655       & 0.599       &        & 0.617       & 0.945       & 0.594       & 0.604       & 0.585       & 0.675       \\ \hline
\textbf{JI} & 0.636       & 0.589       & 0.643       & 0.346       & 0.616       & 0.617       &        & 0.841       & 0.377       & 0.426       & 0.462       & 0.502       \\ \hline
\textbf{LO} & 0.981       & 0.978       & 1.000       & 0.897       & 0.937       & 0.945       & 0.841       &        & 0.883       & 0.892       & 0.884       & 0.870       \\ \hline
\textbf{MG} & 0.562       & 0.627       & 0.608       & 0.402       & 0.574       & 0.594       & 0.377       & 0.883       &        & 0.446       & 0.490       & 0.539       \\ \hline
\textbf{MZ} & 0.573       & 0.645       & 0.639       & 0.481       & 0.639       & 0.604       & 0.426       & 0.892       & 0.446       &        & 0.511       & 0.567       \\ \hline
\textbf{SO} & 0.582       & 0.636       & 0.620       & 0.519       & 0.519       & 0.585       & 0.462       & 0.884       & 0.490       & 0.511       &        & 0.574       \\ \hline
\textbf{TE} & 0.708       & 0.688       & 0.703       & 0.550       & 0.586       & 0.675       & 0.502       & 0.870       & 0.539       & 0.567       & 0.574       &        \\ \hline
\end{tabular}
\caption{LD Matrix, data from Kirk 1966 : 311 nouns.}
\label{table2}
\end{table}

All networks emerging from this wider and more consistent sample confirm previous results: at $d = 0.43$ (see Fig. \ref{fig6} (Upper Left), we find again two choremes -- one located in the Southern Lowlands, i.e. [JA-IX], and another located in the Central Highlands, i.e. [HU-JI-MG]. The latter choreme, though makes up a chain with a very interesting dialect, which was already viewed as ambivalent by Gudschinsky: MZ clusters with [HU-JI-MG] in a [MZ[HU-JI-MG]] chain.

%
%


The main difference with previous clusters at this stage lays in the boldness of aggregates: MZ would be expected to cluster at a later stage of structural identification with the Highlands choreme, and JA should rather cluster first with DO, instead of telescoping IX.
This behavior of the diasystem  is due to the lesser complexity of the data, as suggested above, when analyzing phonological variables in Table \ref{table2}: the simpler the morphological patterns, the more straightforward the results. 
Bolder chains in Fig.~\ref{fig6} (Upper Right) give therefore more clear-cut hints at the deep structure of the diasystem. At $d = 0.55$, an overt extensive rhombus appears, crossing the whole area from west to the east, strongly rooted in MZ in the West and SO in the East, with two lateral extensions: TE in the Northwest and AY in the East. One could not dream of a better r\'esum\'e  of most of our previous observations: TE and AY are outstanding actors as pivots, or transitional spots, while MZ, HU and SO had already been noted as crucial innovative dialects, since the early phases of Gudschinsk's models of differentiation -- stages (3) and (4) in (II) and stage (3a)in (III). At $d = 0.72$, a trapezoid resorting more to a parallelogram than to an isosceles shows up,  confirming the far reaching links between TE and IX, going all the way down towards AY and CQ to climb up toward MZ and reaching TE in a loop -- this geometry actually comprehends the periphery of the diasystem, and may point at a deeper level of structuration.

%
\begin{figure}[ht!]
\centering
\includegraphics[width=0.48\linewidth]{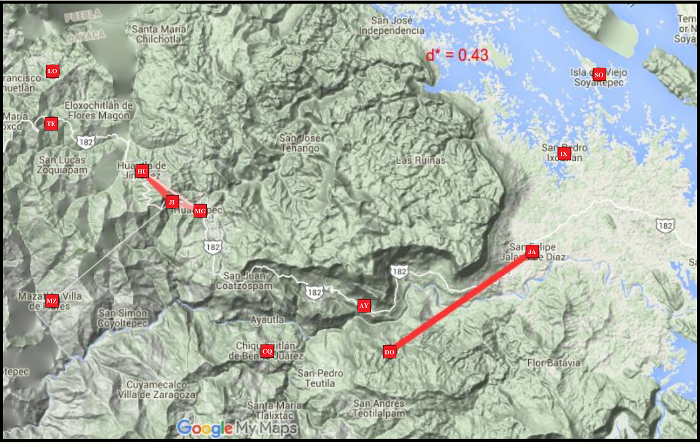}
\includegraphics[width=0.48\linewidth]{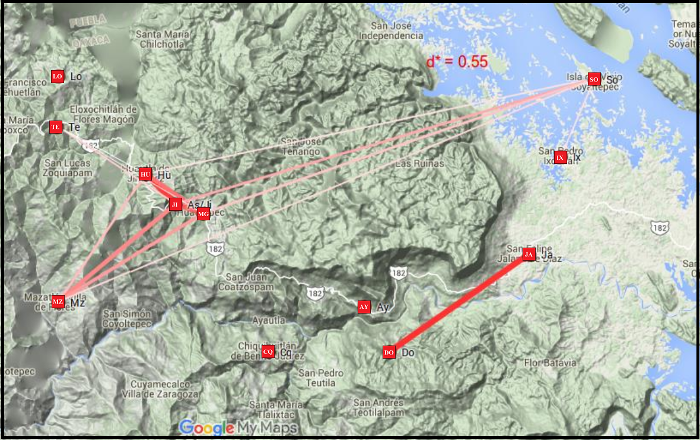}
\includegraphics[width=0.48\linewidth]{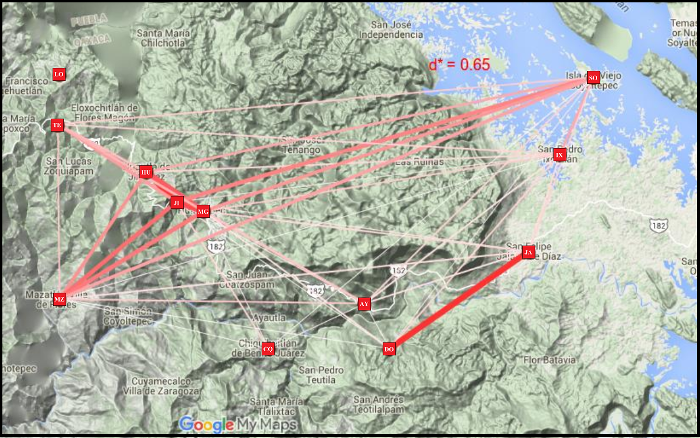}
\includegraphics[width=0.48\linewidth]{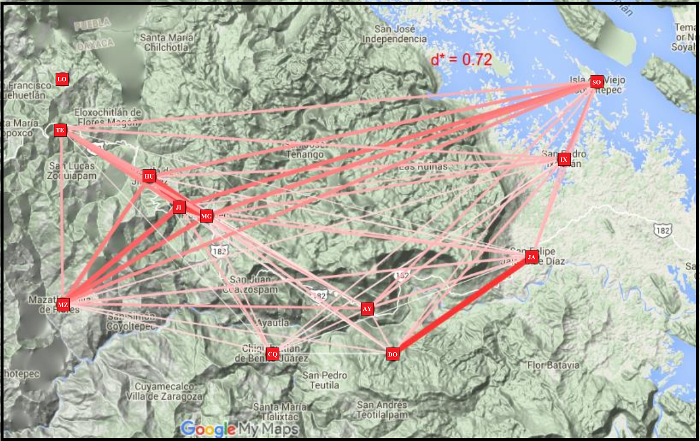}
\caption{LD applied to nouns in Kirk's data. Four thresholds $d = 0.43$ (Upper Left), $0.55$ (Upper Right), $0.65$ (Lower Left) and $0.72$ (Lower Right) of normalized mean distance.}
\label{fig6}
\end{figure}

The Minimum Spanning Tree (MST) diagram in Fig.~\ref{fig7} endows the Central Highlands dialect JI with enhanced centrality. The fact that the transitional variety of AY in the Midlands is intertwined with another ``buffer zone'' dialect, according to Gudschinsky's model, confirms details of the deep structure of the dialect network.

%
\begin{figure}[ht!]
\centering
\includegraphics[width=\linewidth]{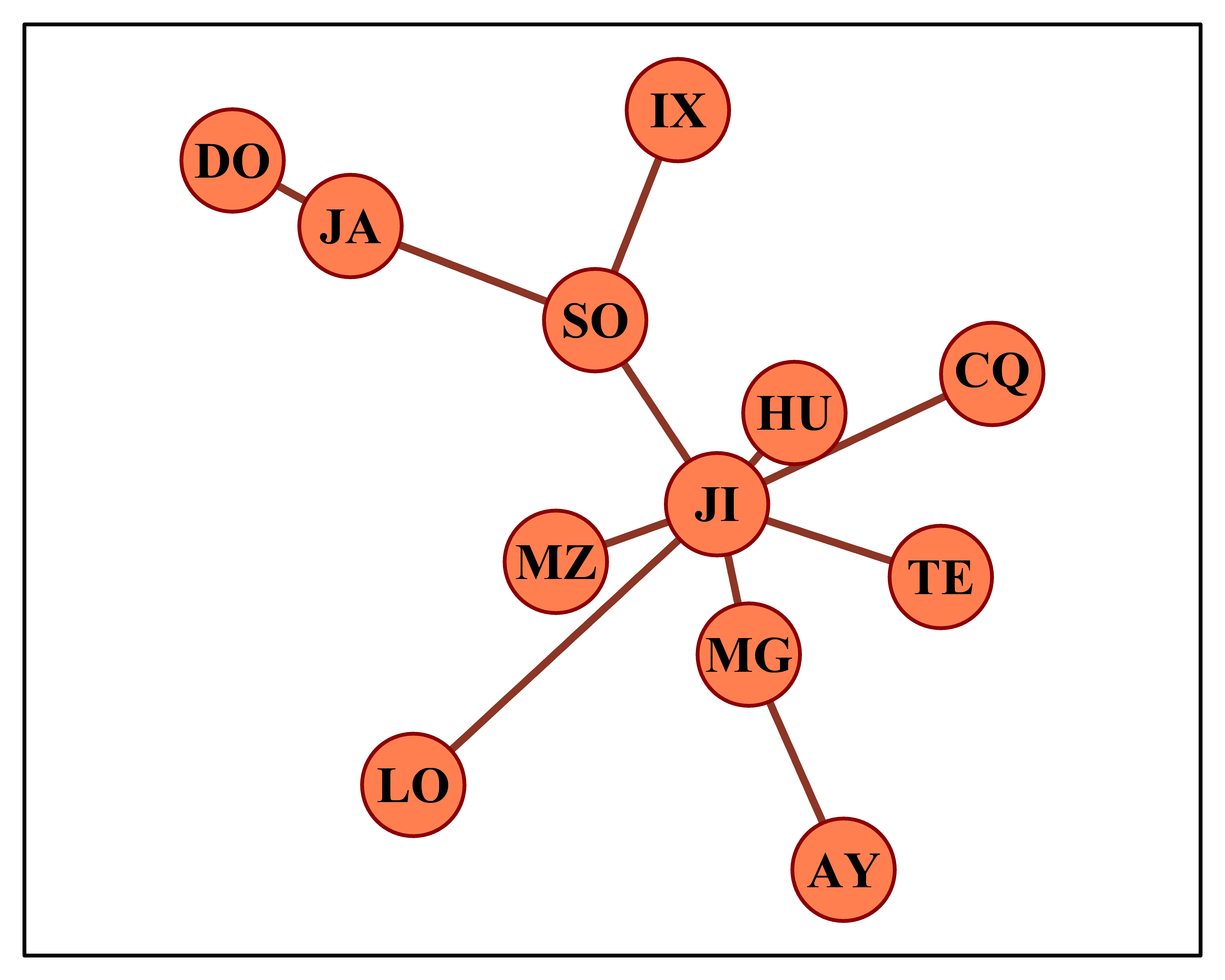}
\caption{Minimum spanning tree based on the LD applied to nouns in Kirk's data.}
\label{fig7}
\end{figure}

A minimum spanning tree is a spanning tree of a connected, undirected graph such that all the $N$ (here $N = 12$) dialects are connected together with the minimal total weighting for its $N-1$ edges (total distance is minimum).
The distance matrix defined by the LDs among the dialects was used as an input to the inbuilt MST function in R (See R documentation for details).
Here, we state Kruskal and Prim algorithms for the sake of completeness of the present article. 

Description of the two algorithms :

\begin{itemize}

\item Kruskal --  {\it This algorithm extends the minimum spanning tree by one edge at every discrete time interval by finding an edge which links two separate trees in a spreading forest of growing minimum spanning trees}. 

\item Prim -- {\it This algorithm extends the minimum spanning tree by one edge at every discrete time interval by adding a minimal edge which links a node in the growing minimum spanning tree with one other remaining node}.\\
 Here, we have used Prim's algorithm to generate a minimum spanning tree.

\end{itemize}

The dendrogram in Fig.~\ref{fig8} does not only provide an overall picture of the dialect network: it tells us more about the intricacy of communal aggregates and layers of differentiation. It also solves a few problems raised by discrepancies between model (II) and (III) and our results. In this Stammbaum, Highlands dialects actually cluster with Lowlands dialects, while Southern Midlands dialects cluster together with a ``default'' variety -- CQ, a near neighbor in the South. In the inner cluster of the dendrogram including Highlands dialects, we come across the [MZ[HU-JI-MG]] chain we are already familiar with, on the one hand, a quite heterogeneous sub-cluster made up of a [IX-SO] chain, associated to the far distant TE Northwestern Highlands dialect, usually classified within the Highland dialects.
Last, but not least, the LO dialect, though we can consider it as a byproduct of a recent Northwestern dialect over differentiation (i.e. from TE), stands on its own, as if it would classify as a totally different language -- which it is not, although its differences are indeed phonologically conspicuous, because of  recent vowel shifts i $\to$ e, e $\to$ a, a $\to$ o, u $\to$ \"i.

%
\begin{figure}[ht!]
\centering
\includegraphics[width=\linewidth]{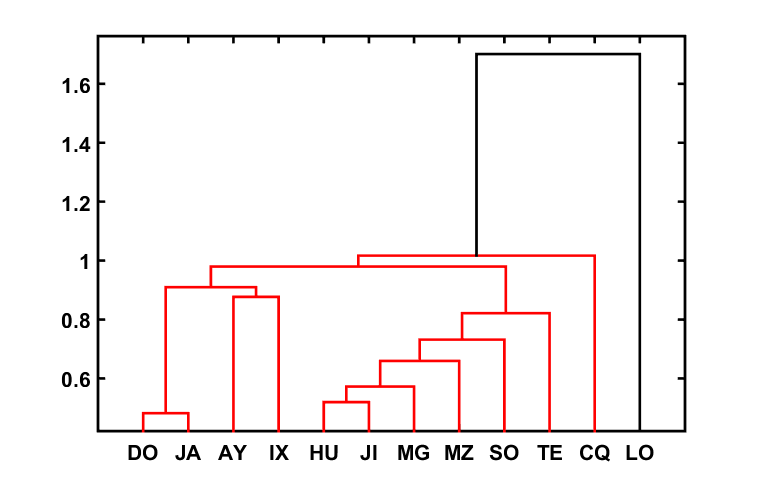}
\caption{LD applied to nouns in Kirk's data: Dendrogram.}
\label{fig8}
\end{figure}

A dendrogram is basically a tree diagram. This is often used to depict the arrangement of multiple nodes through hierarchical clustering. We have used the inbuilt function in MATLAB (see MATLAB documentation) to generate the hierarchical binary cluster tree (dendrogram) of 12 dialects connected by many U-shaped lines (as shown in Fig.~\ref{fig8}), such that the height of each U represents the distance (given by the LDs) between the two dialects being connected. Thus, the vertical axis of the tree captures the similarity between different clusters whereas the horizontal axis represents the identity of the objects and clusters. Each joining (fusion) of two clusters is represented on the graph by the splitting of a vertical line into two vertical lines. The vertical position of the split, shown by the short horizontal bar, gives the distance (similarity) between the two clusters. We set the property ``Linkage Type" as ``Ward's Minimum Variance", which requires the Distance Method to be Euclidean which results in group formation such that the pooled within-group sum of squares would be minimized. In other words, at every iteration, two clusters in the tree are connected such that it results in the least possible increment in the relevant quantity, i.e., pooled within-group sum of squares.

In spite of these discrepancies with expected taxon, the main lesson of this dendrogram lays in the tripartition [Midlands[Highlands-Lowlands]], and the confirmation of the  [MZ[HU-JI-MG]] chain. 
In Fig.~\ref{fig9}, the two-dimensional projection from Multi-Dimensional Scaling (MDS) analysis mends up the formal oddities (we already mentioned), i.e. TE clustering so far from HU, and CQ so close to AY. This representation, obtained with the same data, is far more congruent with standard taxonomy of Mazatec dialects, as in (I) above: it displays a constellation of choremes as [DO-JA] and [JI-HU], and more loosely tightened chains such as [AY[IX]], [MZ[MG[TE]]] and a fairly distant chain [CQ[SO]]. LO, again, stands far apart, as a strongly innovative dialect as far as phonology is concerned -- with strong consequences on morphology too.

%
\begin{figure}[ht!]
\centering
\includegraphics[width=\linewidth]{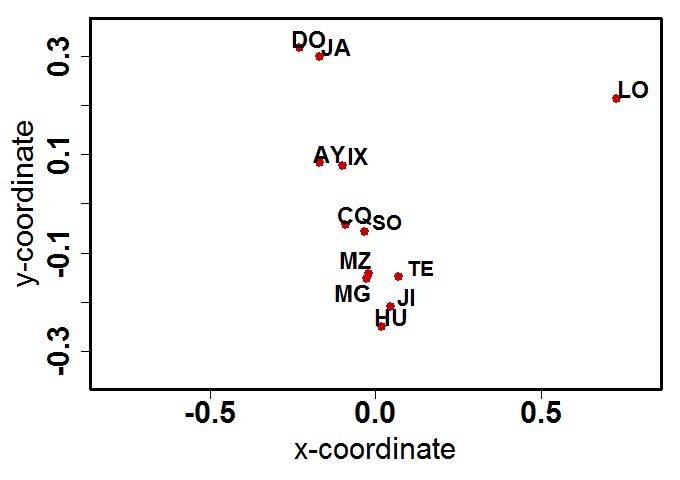}
\caption{Two-dimensional projection from multi-dimensional scaling analysis (in linguistic space). Nouns in Kirk's data.}
\label{fig9}
\end{figure}

MDS is a method to analyze large scale data that displays the structure of similarity in terms of distances, obtained using LD algorithm, as a geometrical picture or map, where each dialect corresponds to a set of coordinates in a multi-dimensional space. MDS arranges different dialects in this space according to the strength of the pairwise distances between dialects -- two similar dialects are represented by two set of coordinates that are close to each other and two dialects behaving differently are placed far apart in space (see Borg 2005). We construct a distance matrix consisting of $N \times N$ entries from the $N$ time series available, defined using LD. Given $D$, the aim of MDS is to generate $N$ vectors $x_1,...,x_N \in \Re^D$, such that
\begin{eqnarray}
\Arrowvert x_i  -  x_j \Arrowvert \approx d_{ij} \hspace{0.3in}\forall i, j\in N, 
\end{eqnarray}
where $\Arrowvert . \Arrowvert$ represents vector norm. We can use the Euclidean distance metric as is done in the classical MDS. Effectively, through MDS we try to  find a mathematical embedding of the $N$ objects into $\Re^D$ by preserving distances. In  general, we choose the embedding dimension $D$ to be  $2$, so that we are able to plot the vectors $x_i$ in the form of a map representing $N$ dialects. It may be noted that $x_i$ are not necessarily unique under the assumption of the Euclidean metric, as we can arbitrarily translate and rotate them, as long as such transformations leave the distances $\Arrowvert x_i - x_j \Arrowvert$ unaffected. Generally, MDS can be obtained through an optimization procedure, where $(x_1,...,x_N)$  is the solution of the problem of minimization of a cost function, such as
\begin{eqnarray}
\min_{x_1,...,x_N} \sum_{i<j} (\Arrowvert x_i - x_j \Arrowvert - d_{ij})^2.
\end{eqnarray}

In order to capture the similarity among the dialects visually, we have generated the MDS plot of 12 dialects. As before, using the International Phonetic alphabets from the database as an input, we computed the distance matrix using the LD algorithm. The distance matrix was then used as an input to the inbuilt MDS function in R.
The output of the MDS were the sets of coordinates, which were plotted as the MDS map as shown in Fig.~\ref{fig9}.
The coordinates are plotted in a manner such that the centroid of the map coincides with the origin $(0,0)$. 

\section{Conclusion and prospects}
\label{conclusion}

As Nicola\"i \& Ploog put it (Nicolaï and Ploog 2013: 278), one has to consider two types of categories, when tackling anything which looks like -- or is supposed to work as -- frontiers: on the one hand, matter or materiality, on the other hand constructs. Matters or materialities rank as follows: geography, geology, biology, ecology, and they partly shape the world we live in, as we are indeed a very adaptive species. Constructs, instead, should be clearly divided in two: compelling patterns on the one hand, elaborations on the other hand. The former range from social constraints or norms, laws, beliefs and habits to economic systems; the latter from models to reforms, according to the activities developed in communal aggregates, in reaction to the environment and its contradictions. 

In this case, matters do matter a lot, as the Mazatec diasystem is vertically structured, from the Lowlands to the Highlands, and some bigger and older centers or town dialects, as JA, HU, MZ, IX indeed weight more than mere villages or hamlets (as JI, MG, AY, CQ, LO). The fact that SO was so peripheral, and ended up as a village nested on the top of a resilient hill above the Miguel Aleman dam, as the village called Viejo Soyaltepec, has consequences on the evolution of certain components of the Mazatec diasystem. The intrusion and the violent reshaping of the whole ecological and socioeconomic settings since the end of the XIXth century, though mercantile activities, instead,have resorted to elaborative constructs, and these have played a strong role too, in smashing previous compelling patterns of inter-communal solidarity or, on the contrary, enmity. Matter and materialities constantly change in nature, indeed, as biology and geology teach us. But cultural constructs as change even faster, and they may even loop, recede and proceed, in a nonlinear way -- so do diasystems throughout history, and so does the Mazatec diasystem in the first place.

But the higher plateau or level in the realm of constructivism and elaboration has to be sought in our models and methods to gather and proceed data, as we did here, handling Kirk's cognate sets, initially collected in order to make a sketch of comparative phonology. We turned it into something quite unexpected, as alchemists used to dream of turning stones or dust into gold. We saw how quantitative tools designed to measure dialect distance, as the Levenshtein algorithm, can provide clues from a Complexity Theory standpoint. Various data sets and a variegated array of computational methods (multi-layered normalized means, minimum spanning tree, multi-dimensional scaling analysis, etc.) applied on these raw sets of data opened the way to a labyrinth of constructs and representations, which teach us a lot about what mattered, in the past, and what matters and will, today and for the future, in such a strongly diversified communal aggregates that make up the Mazatec small world (L\'eonard \& dell'Aquila 2014). 

A world full of complexity, whose survey with the help of Complexity Theory methods suggest that tree-models (Stammbaum), chain models, choremes and buffer zones or transitional areas are not sufficient to grasp geolinguistic complexity. We also have to resort to concepts as pivots, default varieties, and a few more. Neither is the punctuated equilibrium (Dixon 1997) concept enough, as the Mazatec dialect network geometry shows an intricate web of constant interactions. The valley leading from the Lowlands to the Highlands has not only once in a while served as a bottleneck: it seems to be a highway for diffusion and linguistic change which never rests. Corridors from the Northern Midlands, as Santa Maria Chilchotla, and the San Jos\'e enango area, between HU and San Jos\'e Independencia, may also account for this multisource and multidirectional percolation of change and metatypes between communal aggregates. 
The intricate geometry of diasystems has still to be disentangled, and this Mazatec case study provides but a glimpse at how to tackle this issue. Complexity Theory undoubtedly should be at the forefront of such a crucial endeavor, for the understanding of how complex adaptive and cooperative systems such as language and society work and mingle together.

\section{Abbreviations}
AY = Ayautla, CQ = Chiquihuitl\'an, DO = Santo Domingo, IX = San Pedro Ixcatl\'an, JI = Jiotes (or , HU = Huautla, JA = Jalapa, LO = San Lorenzo, MG = San Miguel Huautla, SMt = San Mateo Yoloxochitl\'an, SO = San Miguel Soyaltepec, TE = San Jerónimo Tecoatl (Abbreviations as in Kirk 1966).

\begin{acknowledgement}
M.P. and E.H. acknowledge support from the Institutional Research Funding IUT (IUT39-1) of the Estonian Ministry of Education and Research.
K.S. thanks the University Grants Commission (Ministry of Human Research Development, Govt. of India) for her senior research fellowship.
A.C. acknowledges financial support from grant number BT/BI/03/004/2003(C) of Government of India, Ministry of Science and Technology, Department of Biotechnology, Bioinformatics Division and University of Potential Excellence-II grant (Project ID-47) of the Jawaharlal Nehru
University, New Delhi, India.
\end{acknowledgement}


%
%

\section*{References}

$~~~~~$-- Anderson , P.W. 1972. More Is Different , Science 177, 393-396.

-- Balev Stefan, Jean L\'eo L\'eonard \& G\'erard Duchamp 2016. ``Competing models for Mazatec Dialect Intelligibility Networks'', in L\'eonard, Jean L\'eo; Didier Demolin \& Karla Janir\'e Avil\'es González (eds.). 2016. Proceedings of the International Workshop on Structural Complexity in Natural Language(s) (SCNL). Paris, 30-31 May 2016: Paris 3 University - Labex EFL (PPC11). Available on http://axe7.labex-efl.org/node/353. 

-- Beijering, K, C. Gooskens \& W. Heeringa 2008. ``Predicting intelligibility and perceived linguistic distance by means of the Levenshtein algorithm'', Amsterdam, Linguistics in the Netherlands, 2008,), p. 13-24. 

-- Bolognesi, R. \& W. Heeringa 2002. ``De invloed van dominante talen op het lexicon en  de fonologie  van Sardische dialecten''. In: D. Bakker, T.  Sanders,  R.  Schoonen  and  Per  van der  Wijst  (eds.). Gramma/TTT:  tijdschrift  voor  taalwetenschap.  Nijmegen University Press, Nijmegen, 9 (1), p. 45-84.

-- Borg, I. and Groenen, P., Modern Multidimensional Scaling: theory and applications (Springer Verlag, New York, 2005).

-- Castellano C., S. Fortunato, V. Loreto, Statistical physics of social dynamics, Rev. Mod. Phys. 81 (2009) 591.

-- Dixon, Robert, M. W. 1997.  The Rise and Fall of Languages, Cambridge, Cambridge University Press.

-- Goebl, Hans 1998. ``On the nature of tension in dialectal networks. A proposal for interdisciplinary research'', in Altmann, Gabriel \& Walter Koch (eds.) Systems. New Paradigms for the Human Sciences, Berlin, Walter de Gruyter: 549-571. 

-- Gudschinsky, Sarah, 1955, ``Lexico-Statistical Skewing from Dialect Borrowing'', IJAL 21(2), 138-149.

-- Gudschinsky Sarah, ``Mazatec dialect history'', Language, n° 34, 1958, p. 469-481.

-- Gudschinsky Sarah, 1959. Proto-Popotecan. A Comparative Study of Popolocan and Mixtecan, IJAL, n° 25-2.

-- Heeringa, W. \& C. Gooskens 2003. ``Norwegian dialects examined perceptually and acoustically'', Computers and the Humanities, 57 3: 293-315. 

-- Heinsalu E., Marco Patriarca, Jean L\'eo L\'eonard, 2014.The role of bilinguals in language competition, Adv. Complex Syst. 17, 1450003.

-- Jamieson Carole, 1996. Diccionario mazateco de Chiquihuitl\'an, Tucson, SIL. 

-- Jamieson Carole, 1988. Gram\'atica mazateca. Mazateco de Chuiquihuitl\'an de Ju\'arez, M\'exico D.F,  SIL.

-- Killion Thomas \&  Javier Urcid  2001. ``The Olmec Legacy: Cultural Continuity and Change in Mexico's Southern Gulf Coast Lowlands'',  
Journal of Field Archaeology, 28 1/2: 3-25

-- Kirk, Paul Livingston 1966. Proto-Mazatec phonology. PhD dissertation, University of Washington.

-- Kirk, Paul Livingston 1970. ``Dialect Intelligibility Testing: The Mazatec Study'', International Journal of American Linguistics, Vol. 36, 3 : 205-211.

-- L\'eonard, Jean L\'eo ; Vittorio dell'Aquila \& Antonella Gaillard-Corvaglia 2012. ``The ALMaz (Atlas Lingüístico Mazateco): from geolinguistic data processing to typological traits'', STUF, Akademie Verlag, 65-1, 78-94.

-- L\'eonard, Jean L\'eo \& Alain Kihm, 2014, ``Mazatec verb inflection: A revisiting of Pike (1948) and a comparison of six dialects'', Patterns in Mesoamerican Morphology, Paris, Michel Houdiard Editeur, p. 26-76. 

-- L\'eonard, Jean L\'eo \& dell'Aquila, Vittorio 2014. ``Mazatec (Popolocan, Eastern Otomanguean) as a Multiplex Sociolinguistic ‘Small World''', in Urmas Bereczki  (ed.). The Languages of Smaller Populations: Risks and Possibilities. Lectures from the Tallinn Conference, 16–17 March, 2012, Tallinn, Ungarian Institute's Series: Miscellanea Hungarica: 27-55.

-- L\'eonard Jean L\'eo 2016. ``Diversification, Diffusion, Contact : Mod\'elisation g\'eolinguistique et complexit\'e''. Lalies, 36, E.N.S. de Paris :  9-79.

-- L\'eonard Jean L\'eo \& Julien Fulcrand 2016. ``Tonal Inflection and dialectal variation in Mazatec'', in Palancar, Enrique \& Jean L\'eo L\'eonard (eds), in Palancar, E. \& L\'eonard, J. L. (eds.) Tone \& Inflection : New Facts and New perspectives, Trends in Linguistics. Studies and Monographs, 296, Mouton de Gruyter: 165-195

-- Meneses Moreno, Ana Bella 2004. Impacto pol\`itico, social y cultural de la presa Miguel Alem\'an en la comunidad mazateca de la isla del viejo soyaltepec, Master Thesis, Mexico, Universidad Autonoma Metropolitana (UAM). 

-- Mufwene, Salikoko S., 2001. The ecology of language evolution, Cambridge, Cambridge University Press.

-- Mufwene, Salikoko, 2012. Complexity perspectives on language, communication, and society, in \'Angels Massip-Bonet \& Albert Bastardas-Boada, Springer Verlag: 197-218.

-- Mufwene, Salikoko S., 2013. ``The ecology of language: some evolutionary perspectives'', in Elza Kioko Nakayama Nenoki do Couto \& al. Da fonologia \'a ecolinguística. Ensaios em homenajem a Hildo Honório do Couto, Brasilia, Thesaurus, pp. 302-327. 

-- Nicola\"i , Robert; Ploog, Katja,  2013. ``Frontières. Question(s) de frontière(s) et frontière(s) en question : des isoglosses à la “mise en signification du monde'', in Simonin, Jacky \& Sylvie Wharton 2013 (eds.). Sociolinguistique du contact. Dictionnaire des termes et des concepts, Lyon, ENS Editions : 263-287. 

-- Patriarca, Marco \& Els Heinsalu 2009. Influence of geography on language competition, Physica A 388: 174.

-- Pike, Kenneth. 1948. Tone Languages. A Technique for Determining the Number and Types of Pitch Contrasts in a Language, with Studies in Tonemic Substitution and Fusion. Ann Arbor: University of Michigan Press. 

-- Ross J. \& A.P. Arkin, 2009. Complex systems: From chemistry to systems biology, PNAS 106, 6433– 6434.

-- San Miguel, M., Eguiluz, V. M., Toral, R. and Klemm, K., 2005. Binary and multivariate stochastic models of consensus formation, Comput. Sci. Eng. 7: 67–73.

-- Sol\'e, R., Corominas-Murtra, B. and Fortuny, J., 2010. Diversity, competition, extinction: The ecophysics of language change, Interface 7: 1647–1664.

-- Stauffer, D. and Schulze, C., 2005. Microscopic and macroscopic simulation of competition between languages, Phys. Life Rev. 2: 89.

-- SSDSH 2011-16. Microrregi\'on 13: Zona Mazateca, Secretaria de Desarrollo Social y Humano (SSDSH).

-- Steels, L. 2011. Modeling the cultural evolution of language, Phys. Life Rev. 8, 339–356.

-- Schwartz, Diana 2016. Transforming the Tropics: Development, Displacement, and Anthropology in the Papaloapan, Mexico, 1940s–1960s, PhD. Dissertation, University of Chicago. 

-- Wichmann, S., 2008.The emerging field of language dynamics, Language and Linguistics Compass 2/3: 442.


\end{document}